\documentclass[conference]{IEEEtran}

\usepackage{float}
\usepackage{mathtools}
\usepackage{amssymb}
\usepackage{url}
\usepackage{tikz} 
\usepackage{fixltx2e}

\usepackage{booktabs} 
\usepackage{array,multirow}
\usepackage{listings} 
\usepackage{balance}
\usepackage[noadjust]{cite}

\usepackage{tikz}
\usepackage{pifont}
\newcommand{\cmark}{\ding{51}}%
\newcommand{\xmark}{\ding{55}}%

\newcommand\diag[4]{%
  \multicolumn{1}{p{#2}|}{\hskip-\tabcolsep
  $\vcenter{\begin{tikzpicture}[baseline=0,anchor=south west,inner sep=#1]
  \path[use as bounding box] (0,0) rectangle (#2+2\tabcolsep,\baselineskip);
  \node[minimum width={#2+2\tabcolsep},minimum height=\baselineskip+\extrarowheight] (box) {};
  \draw (box.north west) -- (box.south east);
  \node[anchor=south west] at (box.south west) {#3};
  \node[anchor=north east] at (box.north east) {#4};
 \end{tikzpicture}}$\hskip-\tabcolsep}}

\newcommand\copyrighttext{%
  \footnotesize Pre-print of the accepted article: Lipcak, J., Rossi, B. (2018) A Large-Scale Study on Source Code Reviewer Recommendation, in 44th Euromicro Conference on Software Engineering and Advanced Applications (SEAA) 2018, IEEE.\\
  \textcopyright 2018 IEEE. Personal use of this material is permitted. Permission from IEEE must be obtained for all other uses, in any current or future media, including reprinting/republishing this material for advertising or promotional purposes, creating new collective works, for resale or redistribution to servers or lists, or reuse of any copyrighted component of this work in other works.
  }
\newcommand\copyrightnotice{%
\begin{tikzpicture}[remember picture,overlay]
\node[anchor=south,yshift=10pt] at (current page.south) {\fbox{\parbox{\dimexpr\textwidth-\fboxsep-\fboxrule\relax}{\copyrighttext}}};
\end{tikzpicture}%
}

\begin{document}
\title{A Large-Scale Study on Source Code Reviewer Recommendation}

\author{\IEEEauthorblockN{Jakub Lip\v{c}\'{a}k and
Bruno Rossi}
\IEEEauthorblockA{Faculty of Informatics\\
Masaryk University,
Brno, Czech Republic\\ Email: jakub.lipcak@gmail.com, brossi@mail.muni.cz}
}

\IEEEoverridecommandlockouts

\maketitle

\copyrightnotice

\begin{abstract}
\textit{Context:} Software code reviews are an important part of the development process, leading to better software quality and reduced overall costs. However, finding appropriate code reviewers is a complex and time-consuming task. \textit{Goals:} In this paper, we propose a large-scale study to compare performance of two main source code reviewer recommendation algorithms (RevFinder and a Na\"ive Bayes-based approach) in identifying the best code reviewers for opened pull requests. \textit{Method:} We mined data from Github and Gerrit repositories, building a large dataset of 51 projects, with more than 293K pull requests analyzed, 180K owners and 157K reviewers. \textit{Results:} Based on the large analysis, we can state that i) no model can be generalized as best for all projects, ii) the usage of a different repository (Gerrit, GitHub) can have impact on the the recommendation results, iii) exploiting sub-projects information available in Gerrit can improve the recommendation results.
\end{abstract}

\begin{IEEEkeywords}
Source Code Reviewer Recommendation, Distributed Software Development, Mining Software Repositories
\end{IEEEkeywords}

\IEEEpeerreviewmaketitle

\section{Introduction}

The source code review process was formalized in 1976 by M. E. Fagan, as a highly structured process based on line-by-line group reviews in the form of \textit{inspections} \cite{IbmFagan}. The code review practices have rapidly changed in recent years \cite{Classification_Scheme}, and the modern code review process has become informal (in contrast to Fagan's definition \cite{IbmFagan}), tool-based \cite{ExpectationsAndOutcomes}, more light-weight, continuous and asynchronous \cite{ConvergentContemporary}.

Software code reviews play nowadays a major role in improving the quality of software in terms of defects identified \cite{ref:kemerer2009impact, Factors_Influencing_Reviews}. However, code reviews are a time-consuming process, with significant amount of human effort involved \cite{UnderstandingBroadcast}. Managers might have concerns that code reviews will slow the projects down \cite{Factors_Influencing_Reviews}. Code reviews could also lead to negative moods in teams if team members fear public criticism caused by others reviewing their source code \cite{SoftwarePeerReviews}.

At a process level, source code review processes in companies nowadays are similar to the processes adopted in open source projects \cite{Factors_Influencing_Reviews,ref:rossi2009analysis,ref:rossi2010modelling}, with a lot of variation in the process steps \cite{Factors_Influencing_Reviews,MotivatorsAndDemotivators}. Furthermore, source code reviews can be expensive: knowledgeable understanding of large code changes by reviewers is a time-consuming process, and finding the most knowledgeable code reviewers for the source code parts to be reviewed can also be very labor-intensive for developers. Thongtanunam et al. \cite{RevFinder} examined comments from more than 1,400 representative review samples of four open source projects. Authors discovered that 4\%-30\% of reviews have a code reviewer assignment problem, taking approximately 12 days longer to approve a code change with a code reviewer assignment problem, increasing projects costs.


The reduction of such costs was the main reason for the emergence of source code reviewer recommendation approaches \cite{OnTheNeed, RevFinder, Correct, CoreDevRec, CommenNetwork, ReviewBot, ImprovingCodeReviews} and for the appearance of automation in software engineering, beneficial also for industry \cite{ref:dedik2016automated,ref:roy2014towards}. The automated recommendation of code reviewers can reduce the efforts necessary to find the best code reviewers, cutting time spent by code reviewers on understanding large code changes \cite{OnTheNeed}. This can improve the effectiveness of the \textit{pull-based model}, with source code reviewers assigned to the pull requests immediately after their creation.

In previous studies about source code reviewers recommendation algorithms, the compared projects were limited in size and number, with typically 8K-45K pull requests considered (~\cite{ref:zanjani2016automatically,CoreDevRec,RevFinder,ref:xia2015should,ref:ouni2016search}). In this paper, we look into a large-scale evaluation (293K pull requests, 51 projects) based on Gerrit and Github repositories, that can give more insights about the performance of reviewers recommendation approaches. Our goal is to study the impact of different repository type used for the source code recommendation. For this reason we chose two main baseline algorithms: RevFinder---heuristic model based on file paths \cite{RevFinder} and a Na\"ive Bayes-based approach (NB)---probabilistic model based on additional features such as owners. 
More advanced algorithms exist, however they require additional features to be built (e.g., collaboration networks between developers), making the mining process more complex. We have two main contributions in this paper:

\begin{itemize}
\item provision of a large dataset of 51 projects mined from  Gerrit (14) and Github (37) with 293,337 total pull requests analyzed, considering 180,111 owners and 157,885 reviewers. The dataset is available on Figshare \cite{Lipcak2018}. 
\item comparison of the results in the two repositories (Gerrit, GitHub) using both RevFinder and Na\"ive Bayes-based approaches in the context of the 51-projects dataset. 
\end{itemize}
The article is structured as follows. In section II, we propose the background on several algorithms for code reviewer recommendation. In section III we have the experimental study design, with research questions,  context, data analysis and replicability information. In section IV, we answer the research questions with discussions and threats to validity. Section V proposes the related works and section VI the conclusions.

\section{Background}
Over the years, many approaches for the automation of source code reviewers recommendation emerged \cite{RevFinder,Correct,CoreDevRec,CommenNetwork,ReviewBot,ImprovingCodeReviews}. We categorize the most relevant existing recommendation algorithms into four groups based on the main features they process and the main techniques they use to recommend code reviewers: i) Heuristic-based approaches, ii) Machine Learning-based, iii) Social Networks-based, iv) Hybrid approaches.

\subsection{Heuristic-based Approaches}
Traditional recommendation approaches process historical project review data and use heuristic-based algorithms to find the most relevant code reviewers. Main algorithms are ReviewBot \cite{ReviewBot}, RevFinder \cite{RevFinder} (that we will explain later, as part of our empirical analysis), and CORRECT \cite{Correct}.

\textit{ReviewBot} is a technique proposed by Balachandran \cite{ReviewBot}. It is a code reviewer recommendation approach based on the assumption that lines of code changed in the pull request should be reviewed by the same code reviewers who had previously discussed or modified the same lines of code (\textit{familiarity assumption}). 





\textit{CORRECT  (Code Reviewer Recommendation based on Cross-project and Technology experience)} is another code reviewers recommendation technique \cite{Correct}. The baseline idea is that if a previous pull request used some similar external library or technology to the current pull request, then the reviewers of the past pull request are also good candidates for the current one (\textit{expertise assumption}).


\subsection{Machine Learning}
Algorithms in this group use different \textit{Machine Learning} techniques for the recommendation of code reviewers. They are mainly different from the previous group, as they first need to build a model based on a training set. One typical approach is by Jeoung et al. \cite{ImprovingCodeReviews}. \label{sec:machine_learning}

\textit{Predicting Reviewers and Acceptance of Patches} is an approach by Jeoung et al. \cite{ImprovingCodeReviews} that uses the Bayesian Network technique to predict reviewers and patch acceptance based on a series of features such as patch meta-data, patch content and bug report information.


\subsection{Social Networks}
Social networks have also been used to determine similarities in communication between developers, suggesting more similar candidates for source code reviews.

The \textit{Comment Network (CN)} approach was proposed by Yu et al. \cite{CommenNetwork} as a code reviewer recommendation approach by analyzing social relations between contributors and developers. CN-based recommendation is based on the idea that the interest of developers can be extracted from their commenting interaction. Developers sharing common interests with the originator of a pull request are considered to be appropriate code reviewers \cite{CommenNetwork}. 


\subsection{Hybrid Approaches}
Algorithms in this group use different approaches (machine learning, social network analysis) for the recommendation of code reviewers. An example is CoreDevRec \cite{CoreDevRec}. \label{sec:hybrid}

\textit{Automatic Core Member Recommendation for Contribution Evaluation (CoreDevRec)} builds a prediction model from historical pull requests using file paths similarities, developers GitHub relationship features, and activeness features, using Support Vector Machine (SVM) for prediction \cite{CoreDevRec}.

Table \ref{tab:features} shows a summary of all the features utilized by the reviewed algorithms, while Table \ref{tab:datasets} summarizes the availability of source code, datasets, and metrics used for empirical validation of the proposed algorithms.

\begin{table}[!htb]
  \caption{Summary of features used in most common source code reviewers recommendation algorithms. v=used, x=not used}
  \label{tab:features}
\centering
  \begin{tabular}{ | l | c | c | c | c | c | c | c |}
    \hline
    \multirow{2}{*}{{\begin{tabular}[c]{@{}c@{}}\\ \\ \\ \textbf{Feature}\end{tabular} }} & \multicolumn{6}{c|}{\textbf{Paper}} & \\\cline{2-7}
	& \rotatebox[origin=c]{270}{ReviewBot \cite{ReviewBot}} & \rotatebox[origin=c]{270}{RevFinder \cite{RevFinder}} & \rotatebox[origin=c]{270}{CORRECT \cite{Correct}} & \rotatebox[origin=c]{270}{CoreDevRec \cite{CoreDevRec}} & \rotatebox[origin=c]{270}{Pred. Rev. \cite{ImprovingCodeReviews}} & \rotatebox[origin=c]{270}{Comm. Net. \cite{CommenNetwork}}  & 
\multicolumn{1}{c|}{\textbf{\rotatebox[origin=c]{270}{Total}} }

    \\ \hline \hline 
      \textit{File paths} & \xmark & \cmark  & \xmark & \cmark & \xmark & \xmark & 2  \\ \hline
     \textit{Social interactions} & \xmark & \xmark  & \xmark & \cmark & \xmark & \cmark & 2   \\ \hline
    \textit{Line change history} & \cmark & \xmark & \xmark & \xmark & \xmark & \xmark & 1  \\ \hline
    \textit{Reviewer expertise} & \xmark & \xmark  & \cmark  & \xmark & \xmark & \xmark & 1   \\ \hline
    \textit{Activeness of reviewers} & \xmark & \xmark  & \xmark & \cmark & \xmark & \xmark & 1   \\ \hline
    \textit{Patch meta-data} & \xmark & \xmark  & \xmark & \xmark & \cmark & \xmark & 1   \\ \hline
    \textit{Patch content} & \xmark & \xmark  & \xmark & \xmark & \cmark & \xmark & 1   \\ \hline
    \textit{Bug report information} & \xmark & \xmark  & \xmark & \xmark & \cmark & \xmark & 1   \\ \hline
    \textbf{Total} & 1 & 1  & 1 & 3 & 3 & 1 & \diag{.1em}{.5cm}{}{} \\\hline
  \end{tabular}
\end{table}

\setlength\tabcolsep{4.0pt} 
\begin{table}[htbp]
\caption{Source code \& datasets availability, metrics used in the empirical studies.v=used, x=not used}
\small
\centering
  \begin{tabular}{ | l | c | c | c | c | c | c | c |}
    \hline
    \multirow{2}{*}{{\begin{tabular}[c]{@{}c@{}}\\ \\Paper\end{tabular} }} & \multirow{2}{*} {\textbf{\begin{tabular}[c]{@{}c@{}}\\Source code\\available\end{tabular} }} & \multirow{2}{*} {\textbf{\begin{tabular}[c]{@{}c@{}}\\Dataset\\available\end{tabular} }} & \multicolumn{5}{c|}{\textbf{Metrics}} \\\cline{4-8}
	& & & \rotatebox[origin=c]{270}{ Top-k} & \rotatebox[origin=c]{270}{ MRR} & \rotatebox[origin=c]{270}{ Precision} & \rotatebox[origin=c]{270}{ Recall} & \rotatebox[origin=c]{270}{ F - Measure } \\ \hline \hline 
    \textbf{ReviewBot \cite{ReviewBot}} & \xmark & \cmark & \cmark & \xmark & \xmark & \xmark & \xmark \\ \hline
    \textbf{RevFinder \cite{RevFinder}} & \xmark & \xmark & \cmark & \cmark & \xmark & \xmark & \xmark \\ \hline
    \textbf{CORRECT \cite{Correct}} & \xmark & \xmark & \cmark & \cmark & \cmark & \cmark & \xmark \\ \hline
    \textbf{CoreDevRec \cite{CoreDevRec}} & \xmark & \xmark & \cmark & \cmark & \xmark & \xmark & \xmark \\ \hline
    \textbf{Pred. Rev. \cite{ImprovingCodeReviews}} & \xmark & \xmark & \cmark & \xmark & \xmark & \xmark & \xmark \\ \hline
    \textbf{Comm. Net. \cite{CommenNetwork}} & \xmark & \xmark & \xmark & \xmark & \cmark & \cmark & \cmark \\ \hline
  \end{tabular}
  \label{tab:datasets}
\end{table}
\setlength\tabcolsep{6.0pt} 

\section{Empirical Study Design}
The goal of this paper is to analyze source code recommendation algorithms, with the purpose of investigating i) performance on a large set of projects, ii) the impact of different repository type used for the source code recommendation. We focus on two main algorithms: RevFinder and a novel Na\"ive Bayes-based implementation, as these can give a good baseline for comparison. We have three main research questions:

\begin{itemize}
\item \textbf{RQ1:} \textit{What is the comparison between RevFinder and NB algorithms in terms of performance on both Gerrit and Github Repositories?} In this research question, we investigate two major algorithms for source code recommendation on 51 projects, based on top-k accuracy and Mean Reciprocal Rank (MRR) metrics.
\item \textbf{RQ2:} \textit{What is the impact of using data from GitHub or Gerrit for source code reviewers recommendation?} In this research question, we look at the impact of different repositories and their characteristics in source code recommendation results.
\item \textbf{RQ3:} \textit{Is there any performance improvement when considering the additional sub-projects feature (Gerrit-specific)?} In this research question, we look at exploiting the additional sub-projects feature available in the Gerrit repositories, looking at whether such feature can be useful to improve source code recommendation performance.
\end{itemize}

\subsection{Context}
For the empirical evaluation, we compare two alternative approaches, RevFinder \cite{RevFinder} based on file paths and a Na\"ive Bayes implementation, based on two main features: file paths and owners of reviews.

\subsubsection{RevFinder}
RevFinder \cite{RevFinder} is an approach based on the location of files included in pull requests. The idea of this method is that files located in similar file paths contain similar functionality and therefore should be reviewed by similar code reviewers. 

The first part of RevFinder approach is the \textit{Code Reviewers Ranking Algorithm}. It compares file paths included in a new pull request with all previously reviewed file paths, by using four string comparison techniques (\textit{Longest Common Prefix, Longest Common Suffix, Longest Common Substring} and \textit{Longest Common Subsequence}).

Code reviewer candidates are assigned points in this step. The more similar the file paths are, the higher is the number of points assigned to the code reviewers who previously reviewed them. The second important part of the RevFinder approach is the combination technique. The results of each of the four string comparison techniques are combined using the \textit{Borda count}  combination method. A sorted list of candidates with their scores is then returned as the output of the RevFinder algorithm \cite{RevFinder}. 

We reimplemented the algorithm based on the information that was available in the research paper \cite{RevFinder} and in the GitHub repository for the string comparison part\footnote{\url{https://github.com/patanamon/revfinder}}.

\subsubsection{Na\"ive Bayes Recommendation}
The pseudo-code of our code reviewer recommendation algorithm is shown in Fig. \ref{pseudoCode}. The algorithm takes a new review as input ($R_n$). This review has to contain information about all modified \textit{File paths}, the \textit{Project name} and about the \textit{Owner} of the review. A sorted list of recommended code reviewer candidates ($C$) is returned as the output. The prediction model has to be built at the beginning from all previously closed reviews (lines 5 and 6). 

Lines 8 to 18 describe the main loop where code reviewer candidates are recommended. It iterates over all files modified in the review. Feature probabilities are computed separately for every file and therefore code reviewers are recommended separately. Lines 14 to 17 calculate scores achieved by code reviewers. Every code reviewer is assigned a whole number based on his position in the recommendation list. All the code reviewers who reviewed at least one change request in the past will appear in the recommendation list thanks to \textit{probability smoothing} \cite{ProbabilitySmoothing}. Score calculation is done for every file and the achieved scores are added together. 

Finally, retired reviewers are concatenated to the end of the result list by the \textit{removeRetiredReviewers} function (line 19). This function iterates over all code reviewers and moves down those who have not done any code reviews in recent \textit{n} months. Finally (line 20), code reviewers are sorted by their scores and a sorted list of reviewer candidates is returned as the result. 

\begin{figure} [!htb]
\begin{lstlisting}[mathescape][frame=single]
(*\bfseries Input: *)
$R_n$ : A new review
(*\bfseries Output: *)
$C$: Sorted list of code reviewer candidates
$pastReviews$ $\leftarrow$  List all previously closed reviews 
$bayesRec \leftarrow buildModel(pastReviews)$
$Files_n \leftarrow getFiles(R_n)$
(*\bfseries for *) $fn \in Files_n$ (*\bfseries do *)
    # Get sorted list of recommended reviewers 
    # for every file to be reviewed
    $reviewers \leftarrow bayesRec.recom(f_n, R_n.own, R_n.prjName)$    
    # Assign points to reviewers  
    $scoreCount \leftarrow 0$
    (*\bfseries for *) $r \in reviewers$ (*\bfseries do *)
        $scoreCount++$
        $C[r].score \leftarrow C[r].score + reviewers.len - scoreCount$
    (*\bfseries end for *)
(*\bfseries end for *)
$C \leftarrow removeRetiredReviewers(C)$
(*\bfseries return *)$C.sortBy(score)$
\end{lstlisting}
\caption{Na\"ive Bayes-based \textit{Code Reviewer Recommendation Algorithm}.}
\label{pseudoCode}
\end{figure}

\begin{table*}
  \caption{Mined GitHub and Gerrit Repositories}
  \label{tab:repos}
  \centering
  \begin{tabular}{lccccccccccccc}
    \toprule
    \multirow{2}{*}[-3pt]{Repository}  & \multirow{2}{*}[-3pt]{\# Projects} & \multicolumn{4}{c}{pull requests} & \multicolumn{4}{c}{reviewers} & \multicolumn{4}{c}{owners} \\ 
        \cmidrule{3-6} \cmidrule{7-14} 
      &  & avg & max& min& tot& avg & max& min& tot& avg & max& min& tot \\
      
    \midrule
    GitHub & 37& 4,323 & 29,807 & 431 & 159,953 & 583 & 1,949& 85 & 21,599 & 1,061 &2,970 & 313 & 39,267\\
    Gerrit& 14 & 9,530 & 31,582 & 344 & 133,424 & 204 & 651 & 30 & 4,558 & 325 & 766 & 27 &4,558\\
    \midrule
    \textbf{Overall}& \textbf{51} & \textbf{5,752} & \textbf{31,582} & \textbf{344} & \textbf{293,377} & \textbf{479} & \textbf{1,949} & \textbf{30} & \textbf{24,461} & \textbf{859} & \textbf{1,061} & \textbf{27} & \textbf{43,825} \\
    \bottomrule
  \end{tabular}
\end{table*}

\subsubsection{RevFinder+ and NB+}
As pull requests in the same subproject are often reviewed by similar code reviewers, we added the sub-project name as a feature of RevFinder and NB algorithms for the recommendation process: we named these versions RevFinder+ and NB+. Unfortunately, this information is only available in the Gerrit repository, so we could only apply it in the context of the projects hosted in Gerrit.

\subsection{Approach}

For the experiments, we gathered data from 51 open source software projects through the \textit{Gerrit} and \textit{GitHub} systems APIs (Table \ref{tab:repos}, which provides information about the number of pull requests, reviewers, and owners in the dataset). We chose 14 large and active \textit{Gerrit} projects and 37 \textit{GitHub} projects, which belonged to the most popular \textit{GitHub} repositories according to the gained stars.

\begin{figure}[H]
\includegraphics[width=\linewidth]{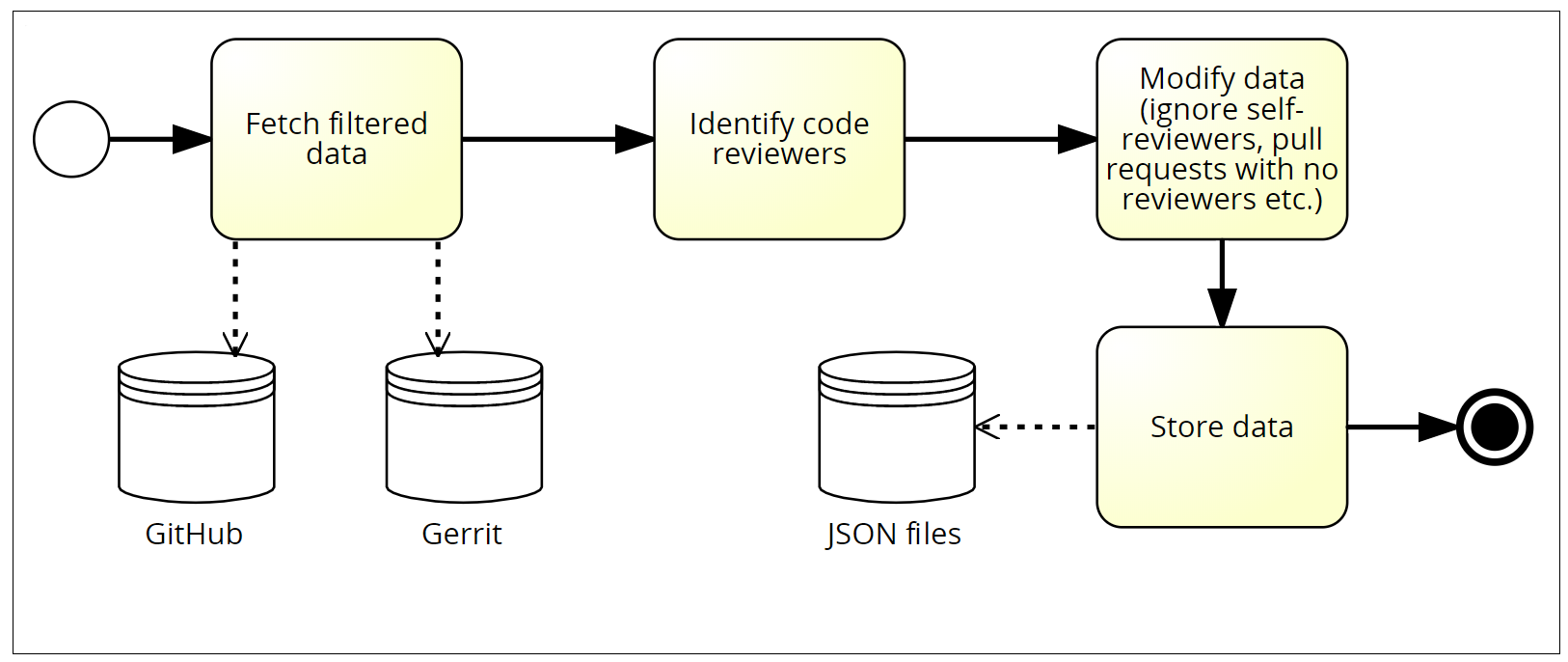}
\caption{Data mining process.}
\label{mining_diagram}
\end{figure}

The REST API provided by the \textit{Gerrit} system was used to collect data from 14 projects involving 133,424 pull requests (Fig. \ref{mining_diagram}, data mining process). Although the API slightly differs between different versions of \textit{Gerrit}, it was always possible to mine the data necessary for our experiments. For each project, we extracted the data from all pull requests with the status 'merged', where all the users who marked the pull request with positive \textit{Code-Review} label were considered as reviewers of the given pull request.

The \textit{GitHub} REST API was used to mine data from 37 projects having 159,953 pull requests in total (Fig. \ref{mining_diagram}, data mining process). We only processed pull requests in a closed state with those users considered as reviewers, who created at least one \textit{Issue Comment} or \textit{Pull Request Review Comment} for the given pull request.

The final dataset contains 293,377 pull requests stored in the form of JSON files. Each JSON file contains the list of pull requests from one project with the following features: \textit{sub-project} (always empty string for \textit{GitHub} projects), \textit{change id}, \textit{change number}, \textit{timestamp}, \textit{owner of the pull request}, \textit{list of reviewers} and \textit{list of modified files}. The owner of a review is the developer that submitted the code review request, while reviewers represent developers that participated in the review process. If the owner of the pull request also appeared as the reviewer of the pull request, he/she was removed from the list of reviewers, because such cases do not need any recommendation. Pull requests with an empty list of reviewers or an empty list of files were removed from the dataset as well.

\textbf{Evaluation Metrics.} Top-k Accuracy, Mean Reciprocal Rank (MRR), Precision and Recall (see also back to Table \ref{tab:datasets}) are used for the evaluation of source code reviewers recommendations. However, we do not use Precision and Recall (and combined F-Measure) as they evaluate the intersection of top-k recommended code reviewers with all actual code reviewers, so they might not be appropriate for datasets containing reviews where the size of the set of actual code reviewers is smaller than the $k$ value \cite{RevFinder}.

\subsubsection{Top-k Accuracy}
The result of \textit{top-k accuracy} is a percentage describing how many code reviewers were correctly recommended within top \textit{k} reviewers of the result list.

\begin{equation}
\textrm{Top-k accuracy} = \frac{\sum\limits_{r \in R} \textrm{isCorrect(r, k)}}{|{R}|} 
\end{equation}
where: \textit{R} is a set of reviews, \textit{isCorrect(r, k)} is a function returning 1 if at least one code reviewer recommended within  top \textit{k} reviewers approved the review \textit{r}, otherwise, 0 is returned.


\subsubsection{Mean Reciprocal Rank}
\textit{Mean Reciprocal Rank (MRR)} is a value describing an average position of actual code reviewer in the list returned by recommendation algorithm.

\begin{equation}
\textrm{MRR} = \frac{1}{|{R}|} \sum\limits_{r \in R} \frac{1}{\textrm{rank(r, recommend(r))}} 
\end{equation}
where: \textit{R} is a set of reviews, \textit{recommend(r)} is a function returning a sorted list of code reviewers recommended for review \textit{r}, \textit{rank(r, l)} is a function returning the rank of the code reviewer who approved review \textit{r} in a sorted list \textit{l}. The value of $\frac{1}{\textrm{rank(r, l)}}$ will be $0$ if there is no such code reviewer in the list \textit{l}.




\subsection{Data Analysis}
\begin{figure}[H]
\includegraphics[width=\linewidth]{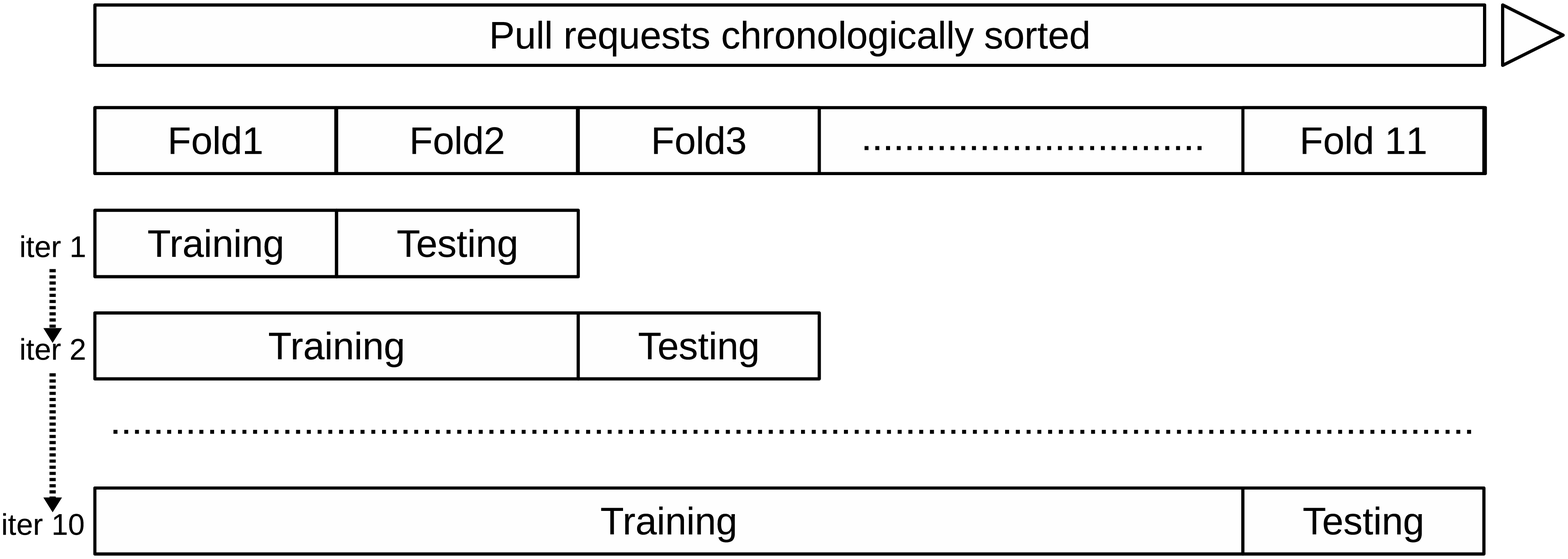}
\caption{Eleven folds experimental set-up}
\label{fig:folds}
\end{figure}

Assignment of code reviewers to pull requests is time dependent. Therefore, we had to use a setup which ensures that only data from past pull requests will be used to recommend code reviewers for future pull requests. For testing the Na\"ive Bayes-based approach we chose 11-fold validation inspired by Bettenburg et al. \cite{foldsValidation} and Jeong et al. \cite{ImprovingCodeReviews}, dividing the training set into 11 equally sized folds. The experiments are run in 10 iterations using different folds as training and test sets (Fig. \ref{fig:folds}). The results of all iterations are then averaged.

RevFinder algorithm always considers all pull requests from the past for recommendation, without needs to build a model for training folds. As we wanted to evaluate the accuracy of both algorithms for the same set of pull requests, we only executed RevFinder recommendation for pull requests in folds 2-11.

To compare differences between groups to answer the research questions, we used different statistical tests. For RQ1 and RQ2 we used the Wilcoxon Signed-Rank Test \cite{ref:rey-wilcoxon2011,ref:kraska-nonparametric2013}, a paired difference test to evaluate the mean ranks differences that we applied to the MRR metric. For RQ3, we used the Mann-Whitney U Test (also Wilcoxon Rank-Sum Test, or just Wilcoxon-Mann-Whitney Test) \cite{ref:macfarland-non-parametr2016,ref:kraska-nonparametric2013} for MRR, due to the fact that our samples for this RQ are not paired, coming from different repositories: Gerrit(14), GitHub (37). We use the more conservative assumption of two-tailed tests, as we do not have any \textit{a priori hypothesis} that one of the compared approaches is better---furthermore, all tests are non-parametric, not assuming normal data distribution \cite{ref:kitchenham-robust2017}. While the $p$-value can show us if the effect is statistically significant, we also report the effect size, that gives an indication of the size of the observed effect, independently from the sample size \cite{ref:cohen-power1992,ref:sullivan-effect-size2012}. For Wilcoxon-Mann-Whitney tests, we calculate effect size as $r=Z / \sqrt{N}, where\: N= \#\, cases*2$, to take into account non-independent paired samples \cite{ref:pallant-spss2013}, using Cohen's definition to discriminate between small ($0.0-0.3$), medium ($0.3-0.6$), and large effects ($>0.6$) \cite{ref:cohen-power1992}.

\textbf{Replicability.} Throughout the study, we followed the concept of reproducible research \cite{ref:madeyski-reproducible2015, ref:madeyski-wider2017}. We implemented a Spring Boot application. We used the \textit{MySql} relational database as a data storage and the \textit{Hibernate} framework for object-relational mapping. We used the \textit{Jayes} library for all the computations involving Na\"ive Bayes.
The dataset is available on Figshare \cite{Lipcak2018}, the source code is available at \url{https://github.com/XLipcak/rev-rec}. 

\section{Empirical Results}
In this section, we discuss the study results based on the three central research questions that were set. Information about all projects considered is available in the appendix Table \ref{tbl:appendix}.

\subsection{RQ1: What is the comparison between RevFinder and NB algorithms in terms of performance on both Gerrit and Github Repositories?}

In this research question, we investigate RevFinder and NB on all 51 projects, based on top-k and MRR metrics.

In terms of average MRR from both the algorithms, results are similar (MRR\textsubscript{RF}=60.6, MRR\textsubscript{NB}=59.1). However, running a paired comparison with a Wilcoxon Signed-Rank Test, paired difference test to evaluate the mean ranks differences for MRR, showed \textbf{significant results} ($p$-value 0.0466. The results are significant at $p\leqslant 0.05$, two-tailed, with \textbf{small effect size} ($r=0.19$)).

\begin{figure} [H] 
\includegraphics[width=\linewidth]{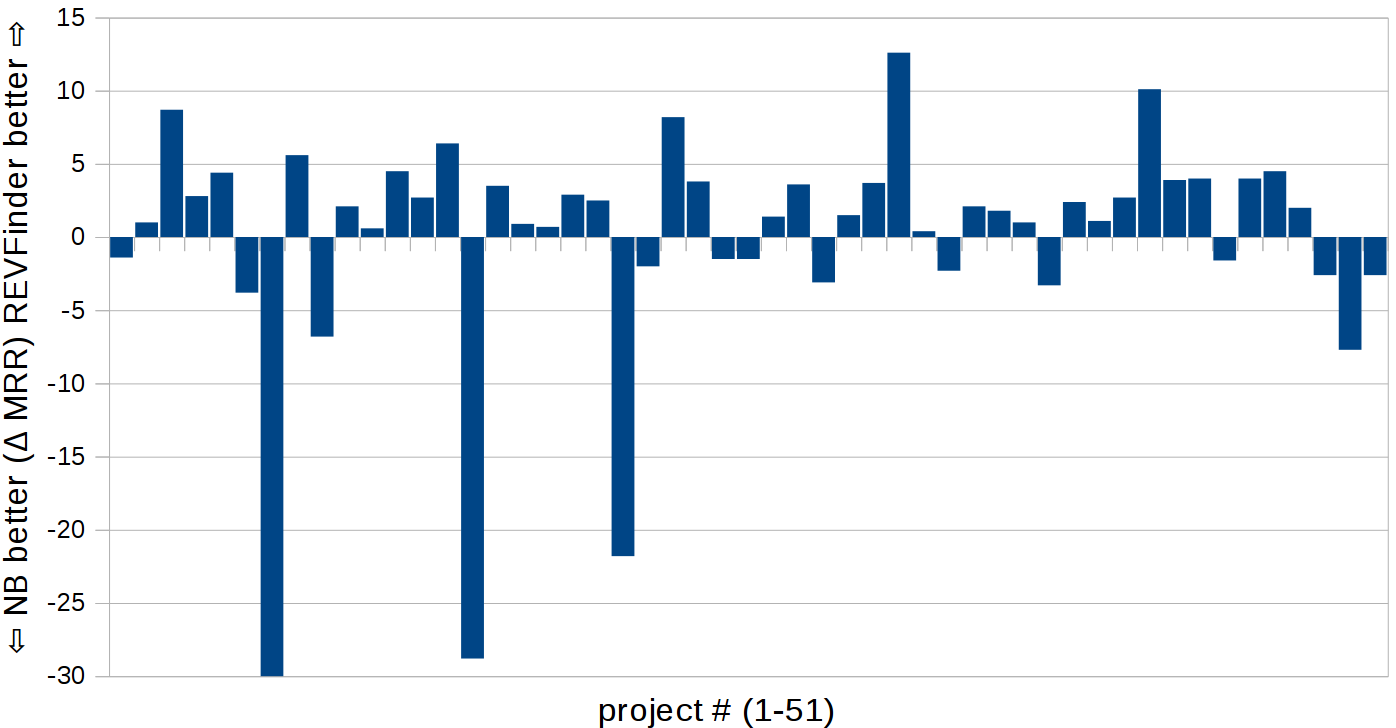}
\caption{RQ1. Delta MRR between RevFinder and NB. Each bar represents one project (positive = gain of RevFinder, negative = gain by NB).}
\label{fig:RQ1-delta}
\end{figure}

Such differences can be seen also when plotting the delta MRR (RevFinder-NB) for each of the 51 projects (Fig. \ref{fig:RQ1-delta}). We observed that for projects Eclipse (+31.1), Lineageos (+28.8) and Qt (+21.8), the NB-approach is consistently better, while RevFinder is, in general, better in the prediction results. We did not find a clear explanation about the difference in the results for these three projects: one possibility is that considering owners for these projects is more relevant than file paths (used in RevFinder) for the choice of the final reviewer. Another explanation is that, as we will discuss in the limitations, all these three projects are from Gerrit, and we could not remove bots in such repository as there is no automated way for identification. So in all these projects, the presence of bots might have made the task of NB easier. Another explanation could be issues in the mining process, Qt was for example removed from the dataset provided in Yang et al. \cite{Yang2016MSR} due to issues in the official Gerrit API.


\textbf{Interpretation of the results.} While the two distributions show similar results, looking at paired differences between predictions on the same projects showed significant results. While generally RevFinder behaves better, three projects showed substantial difference in results in favor of NB---which brings main concerns about the selection of projects in smaller scale comparisons. Effect size is small, meaning that observed differences between RevFinder and NB are small and significance can be due to the sample size used in this study.

\vspace{6pt}
 \par\noindent\fbox{\parbox{\linewidth}{\textbf{Findings.} Within the 51 projects (Gerrit+GitHub), generally RevFinder is better, but effect size is small and differences might be trivial. A simple algorithm based on file paths (RevFinder) without building a model, reaches 60\% MRR.}}
 \vspace{2pt}

\subsection{RQ2: What is the impact of using data from GitHub or Gerrit for source code reviewers recommendation?}
In this research question, we look at the impact of different repositories and their characteristics in source code recommendation results.

We run Wilcoxon U-Test for MRR, top-1, and top-k measures comparing GitHub and Gerrit repositories (Table \ref{tab:bonferroni}). With six comparisons, we applied Bonferroni correction ($\frac{0.05}{6}$), which requires $p \leqslant 0.0083$ to be significant. Results showed \textbf{significant results} for RevFinder for two out of three measures, while they were \textbf{not significant} for NB. In all cases, we can report a \textbf{small effect size}. 

\begin{table}[H]
  \caption{Wilcoxon U-Test performance on GitHub vs Gerrit. **=significant after Bonferroni correction}
  \label{tab:bonferroni}
  \centering
  \begin{tabular}{lll}
    \toprule
     & \multicolumn{1}{c}{RevFinder} & \multicolumn{1}{c}{NB}\\ 
    \midrule
    MRR & $p=0.0071$, $r=0.26$**& $p=0.5619$, $r=0.05$\\
    Top-1 & $p=0.0139$, $r=0.24$ & $p=0.8103$, $r=0.02$\\
    Top-3& $p=0.0080$, $r=0.25$**  & $p=0.3788$, $r=0.08$ \\
    \bottomrule
  \end{tabular}
\end{table}

\begin{figure}[H]
\includegraphics[width=0.95\linewidth]{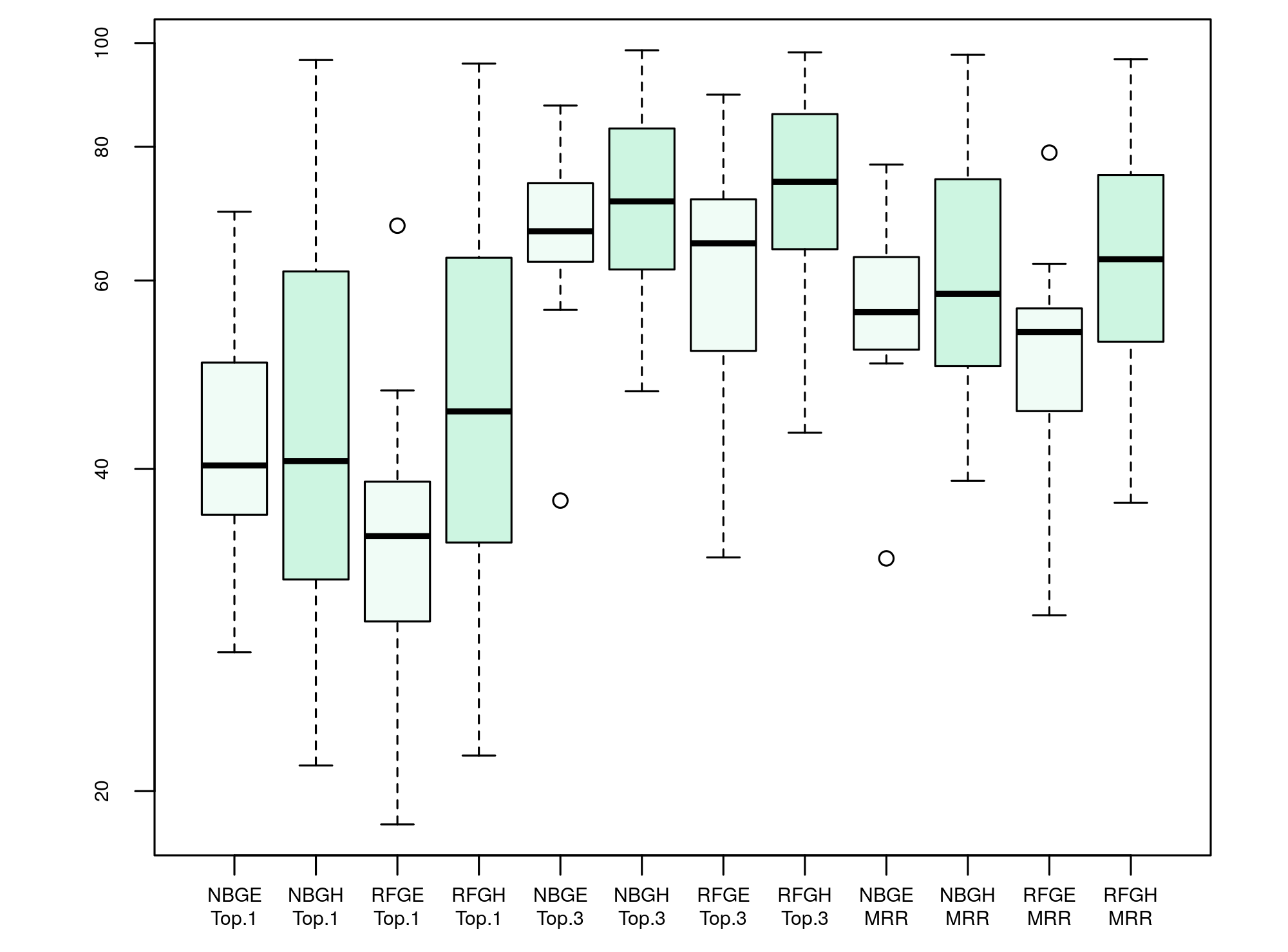}
\caption{RQ2. Comparison of Gerrit and Github datasets for \% top-k and MRR: RF= RevFinder, NB=Na\"ive Bayes, GE= Gerrit, GH=GitHub.}
\label{fig:RQ3}
\end{figure}

If we look at the performance on the different repositories, we notice that results on GitHub are better than those on Gerrit (Fig. \ref{fig:RQ3}). Independently from the algorithm used, just by analyzing a GitHub project can bring better results than in the case of Gerrit (MRR\textsubscript{mean} = 53.87 vs 63.03, which is a quite notable difference). If we look for single algorithms, RevFinder reaches MRR\textsubscript{mean} = 50.41 on Gerrit and 55 on GitHub, while NB has MRR\textsubscript{mean} = 57.32 on Gerrit and 61.62 on GitHub.

\begin{figure}[H]
\includegraphics[width=0.95\linewidth]{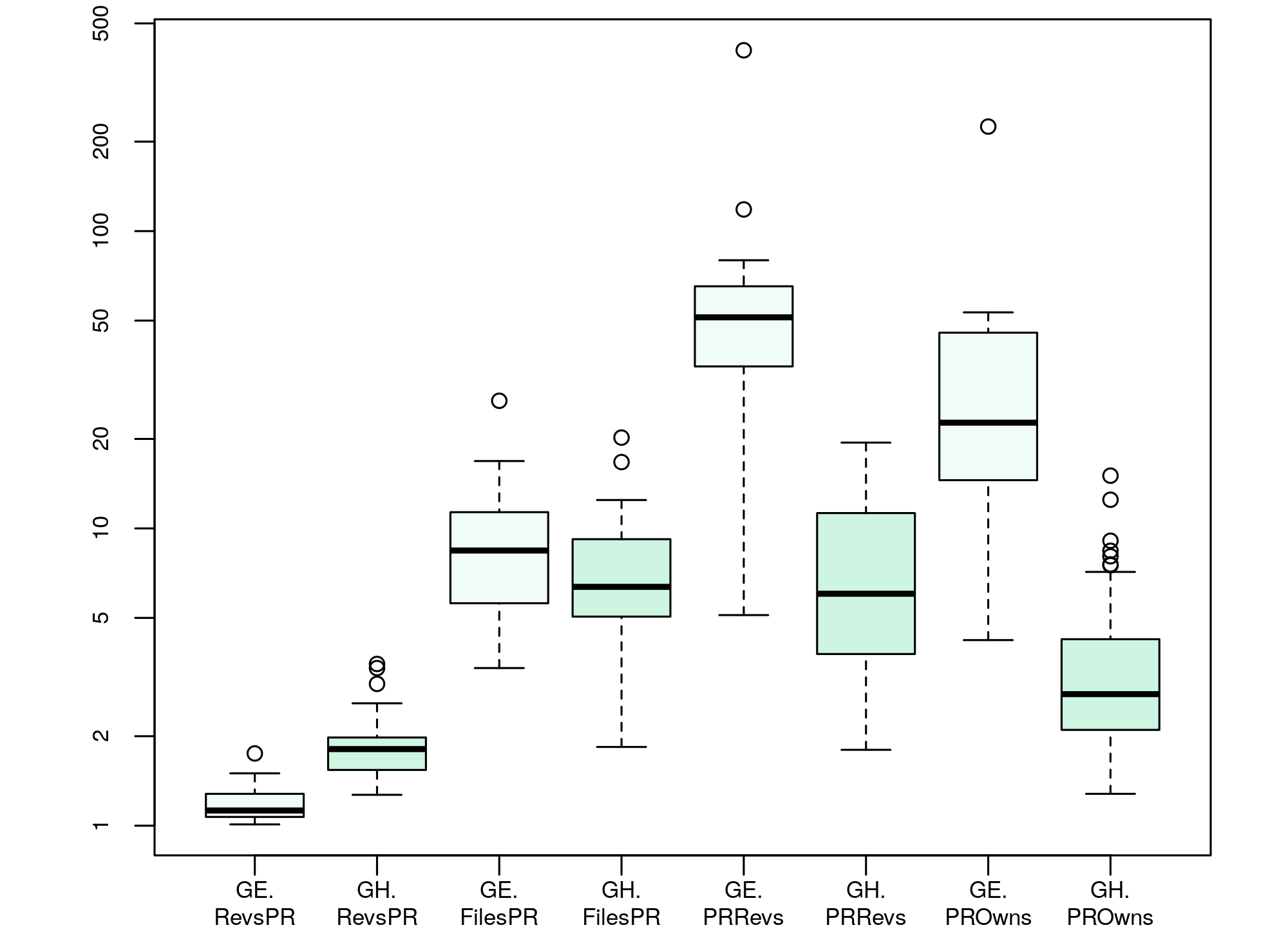}
\caption{RQ2. Comparison of Gerrit and Github datasets (log-scale): GE= Gerrit, GH=GitHub, RevsPR= Reviewers per Pull Request, FilesPR= Files per Pull Request, PRRevs=Pull Requests per Reviewers, PROwns=Pull Requests per Owner.}
\label{fig:RQ3-descriptive}
\end{figure}

\textbf{Interpretation of the results.} As seen, there are differences comparing results obtained on Gerrit or GitHub repositories in terms of recommendation performance. Such variations are statistically significant in the case of RevFinder. Comparisons of different studies need to take this aspect into account, as the application on one repository can get some gains, just for this reason. One interpretation we can give to the results (section \ref{ch:inspection} looks to inspect specific pull requests) is about the differences between the mined repositories. GitHub repositories have a higher number of reviewers per pull request, lower number of files per pull request (Fig. \ref{fig:RQ3-descriptive}). Aspects in which the difference is consistent are a much lower number of pull requests per reviewer and pull requests per owner. Our interpretation is that these factors can have an impact on the final recommendation results. Also in this case, effect size is small, showing that differences can be limited in the practical context.

\vspace{6pt}
 \par\noindent\fbox{\parbox{\linewidth}{\textbf{Findings.} Using different repositories for data analysis can have an impact on the recommendation results in case of the RevFinder algorithm. Running on GitHub can bring better results than on Gerrit projects. However, effect size is small. Our interpretation is that this difference is mainly due to the divergence of the two communities: more general GitHub, more specialized Gerrit.}}
 \vspace{2pt}

\subsection{RQ3: Is there any performance improvement when considering the additional sub-projects feature (Gerrit-specific)?}
In this research question, we look at exploiting the sub-project information feature available in the Gerrit repository.

\begin{figure}[H]
\includegraphics[width=\linewidth]{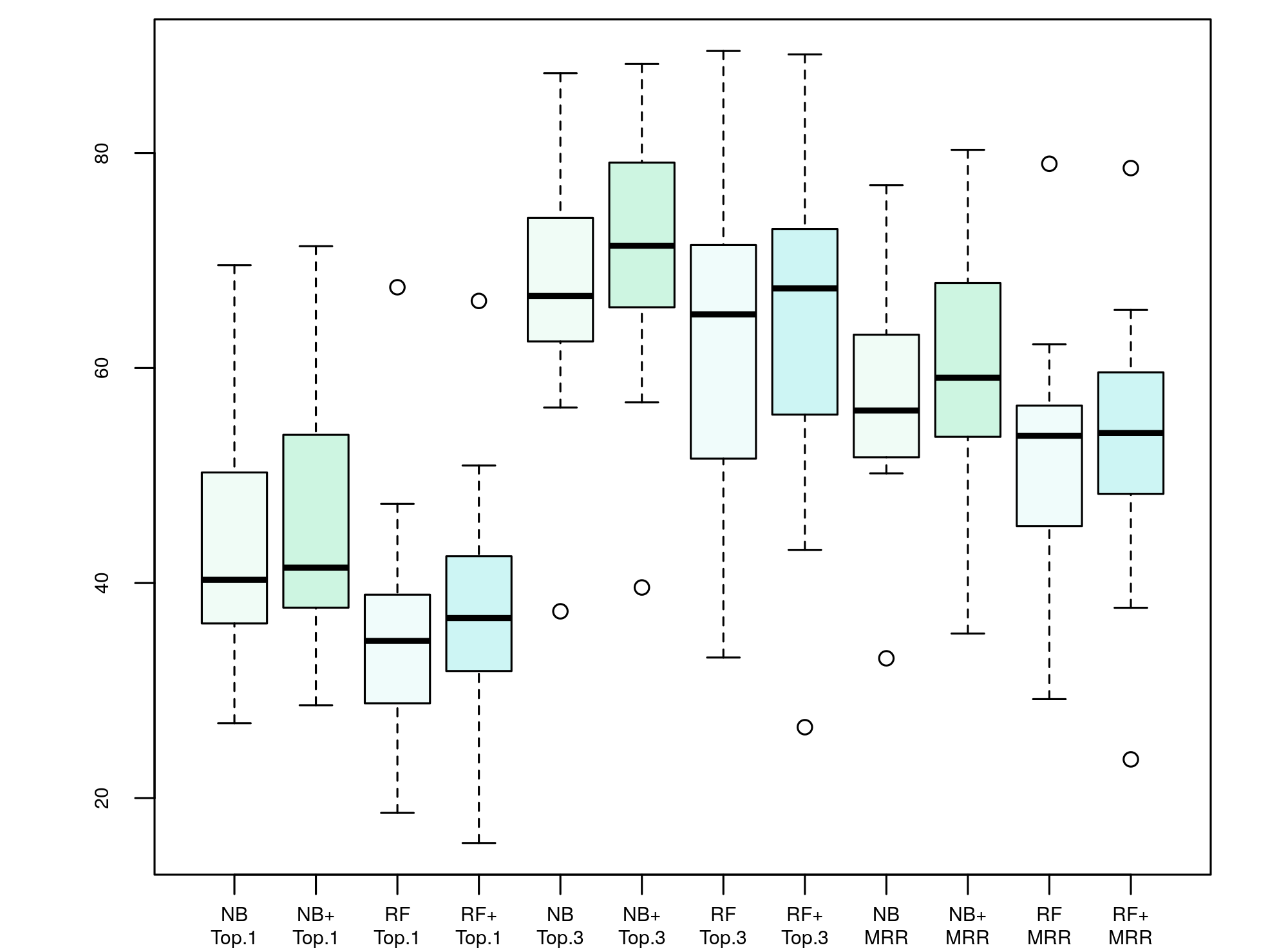}
\caption{RQ3. Using Gerrit additional features for recommendation (RevFinder vs RevFinder+, NB vs NB+).}
\label{fig:RQ2}
\end{figure}

We show MRR and top-k performance for both RevFinder, RevFinder+, NB and NB+. The analysis is run on the 14 Gerrit projects (Fig. \ref{fig:RQ2}).  Running a Wilcoxon Signed-Rank Test, a paired difference test to evaluate the mean ranks differences for MRR results, showed \textbf{non-significant results} for RevFinder+ improvements ($p$-value 0.167. The results are not significant at $p_{bonf}\leqslant 0.025$ ($\frac{0.05}{2}$), two-tailed, with \textbf{small effect size} ($r=0.26$)). The same test for NB+ improvements was instead \textbf{statistically significant} ($p$-value 0.004. The results are significant at $p_{bonf}\leqslant 0.025$ ($\frac{0.05}{2}$), two-tailed, with \textbf{medium effect size} ($r=0.53$)). 

\begin{figure}[H]
\includegraphics[width=\linewidth]{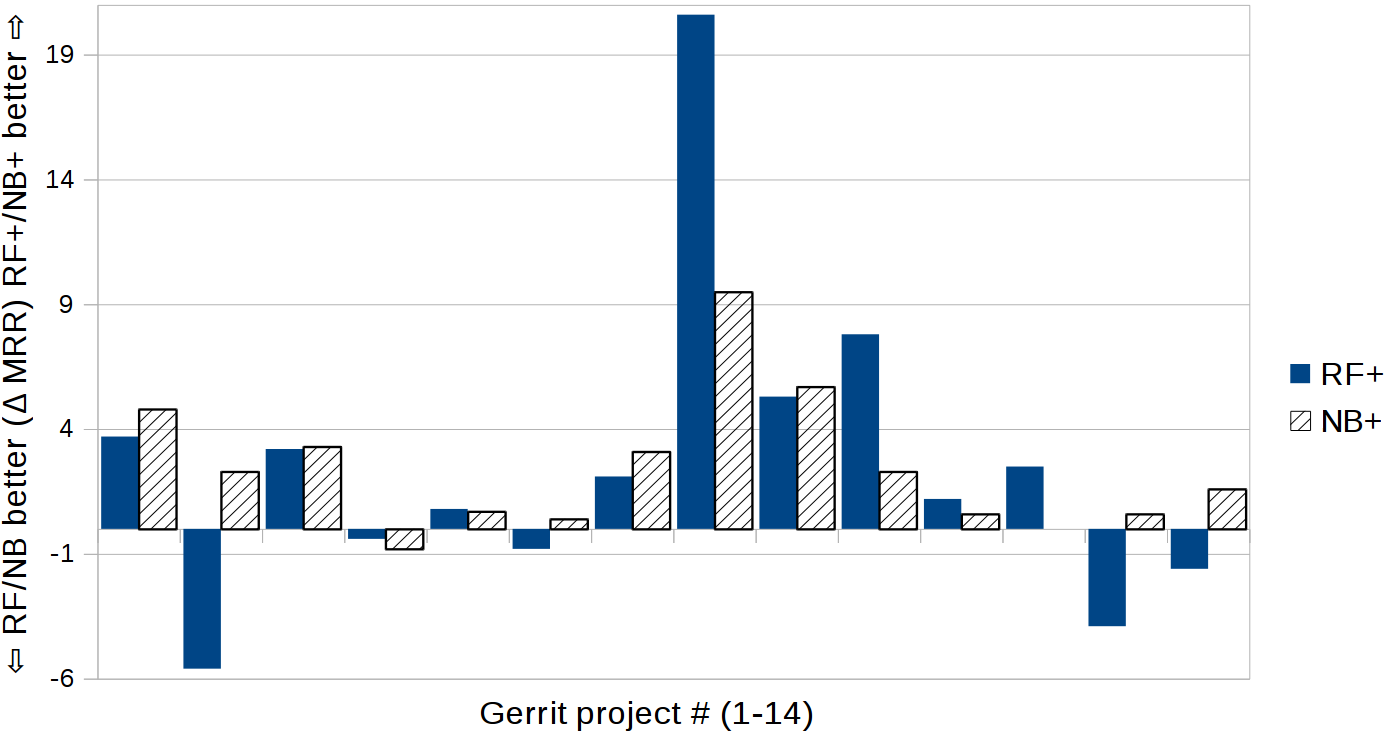}
\caption{RQ3. Delta MRR between RevFinder+ vs RevFinder and NB+ vs NB}
\label{fig:RQ2-delta}
\end{figure}

\textbf{Interpretation of the results.} There is only one project in which NB+ worsens the results slightly compared to NB: in all other projects, considering the sub-projects information with NB improves the outcomes (Fig. \ref{fig:RQ2-delta}). The main suggestion that derives from this analysis is that large distributed projects that are hosted on repositories that support sub-project information can benefit from recommendation automation by using this as a feature. As source code reviewers recommendation algorithms are more useful in case of large distributed system development, such information should be available for improvement of the recommendation results. Medium effect size shows that the differences can be quite noticeable.

\vspace{6pt}
 \par\noindent\fbox{\parbox{\linewidth}{\textbf{Findings.} Considering the 14 Gerrit projects, sub-project information brings significant improvements to the recommendation results for the NB-based approach, with medium effect size.}}
 \vspace{2pt}

\subsection{Evaluation by Inspection} \label{ch:inspection}
To understand the differences between the review process in \textit{Gerrit} and \textit{GitHub}, we also manually checked several pull requests to get an overall sense of the whole process. The most significant differences can be seen in the number of people involved in the review and development process of projects hosted in these systems (Table~\ref{tab:repos}). Although the average size of our datasets mined from \textit{GitHub} is lower than the size of \textit{Gerrit} datasets, the average number of reviews, reviewers and owners is much higher for projects mined from \textit{GitHub}. There are usually fewer people involved in \textit{Gerrit} projects, but these people tend to be more active. Whereas an average reviewer of our \textit{GitHub} dataset reviewed only 7.36 pull requests and an average owner created 3.68 pull requests, 74.91 pull requests were reviewed and 41.1 pull requests were created by an average reviewer/owner in \textit{Gerrit}.

The pull request number 2810\footnote{\url{https://github.com/vuejs/vue/pull/2810}} shows the typical example of \textit{GitHub} review process. User \textit{tejitak} found some issue in the \textit{vue} project and fixed it. He created a new pull request with his solution for this project, which was then discussed by two other community members and was finally merged to the \textit{vuejs:next} branch. It was his first and only pull request in \textit{vue} project and we were able to find many similar cases.

On the other hand, pull request number 71010\footnote{\url{https://go-review.googlesource.com/c/go/+/71010}} shows an example of the typical review process in the \textit{Gerrit} system. It was created for the project \textit{go} by the user \textit{Rob Pike}, who has many other contributions to this project. His pull request was reviewed by the user \textit{Emmanuel Odeke}, who has also already created or reviewed many other pull requests in this project. Such cases are more common in projects hosted by \textit{Gerrit}, where each project is hosted on a different server with a much smaller and more specialized community compared to \textit{GitHub}.

\subsection{Threats to Validity}

\textbf{External validity:} Threats to external validity are related to the generalizability of our study \cite{threats,ref:runeson-guidelines2009}. Our empirical evaluation was executed on 51 open source projects, which use both \textit{Gerrit} and \textit{GitHub} systems for code reviews. We cannot claim that the same results would be achieved for other projects. Smaller open source projects, commercial projects or projects using other systems for code reviews might show different results.

\textbf{Internal validity:} Threats to internal validity are related to experimental errors and biases \cite{threats,ref:runeson-guidelines2009}. The main internal validity threats are about the implementation and the data collection process. We manually inspected results that looked as outliers to fix eventual issues.  The dataset only contains pull requests with at least one reviewer. About bots, no steps were done to reduce them in Gerrit, as there is no automated way for identification. In GitHub, bots usually contain the string "bot" in their name, so those were automatically removed while data mining. The RevFinder implementation was based on the information available in the article and software repository of Thongtanunam et al. \cite{RevFinder}. Still, we might have introduced unintentionally some slight variation. However, experimental results were consistent with previous research results.

A limitation of the \textit{GitHub} REST API that should be taken into account is that responses include a maximum of 300 files per pull request. Pull requests with more than 300 pull requests appear rarely and thus we consider this as a minor limitation. Furthermore, we considered active only developers that had at least one review performed in the last 12 months.

\textbf{Construct validity:} Threats to construct validity relate to how much the measures used represent what researchers aim to investigate \cite{ref:runeson-guidelines2009,ref:wohlin-experimentation2012}.

Our threats are mainly related to the suitability of our evaluation metrics \cite{threats}. We used \textit{top-k accuracy} and \textit{MRR} for empirical evaluation, which are widely used metrics in the relevant literature and should not cause threats to construct validity \cite{RevFinder,Correct,CoreDevRec}. In our experiments, we evaluate whether one of the top-k recommended reviewers actually evaluated the pull request. However, we do not know whether this reviewer was the best candidate for the specific task. Historical data might include wrong reviewers choices.

Source code reviewer recommendation algorithms are primarily expected to be used for large projects where code reviewer assignment problems are more likely to be present \cite{CoreDevRec}, utilization of sub-project information is consistent with this view of larger distributed systems. In a real environment, we would also have to consider some other aspects as well. The most active code reviewers would be probably assigned to a considerable number of pull requests which could lead to overburdening of certain reviewers. Therefore, workload balancing should be also taken into account \cite{RevFinder}.

\section{Related Works}

In this work, we focus on automation of source code recommendation algorithms (\cite{OnTheNeed, RevFinder, ReviewBot, CoreDevRec}), with models often based on the knowledge gathered from previous empirical studies.

The scale of our review is larger than previous studies, in which the maximum number of pull requests considered were ten-times lower orders of magnitude (293K in our case, versus 8K \cite{ref:zanjani2016automatically}, 18K \cite{CoreDevRec}, 42K \cite{RevFinder}, 42K \cite{ref:xia2015should}, 45K \cite{ref:ouni2016search}). Only comparable study to ours is Yu et al. \cite{CommenNetwork}, in which 84 GitHub projects were analyzed, a subset of 650K pull-requests (919 projects, based on the GHTorrent project \cite{ref:GHTorrent2014}).

Due to the scale of the dataset, a comparison with related works concerning MRR and top-k measures is appealing, to see the how the results in previous papers are generalizable. In Thongtanunam et al. \cite{RevFinder}, RevFinder was able to reach top-10 recommendation of 79\% on four projects (Android, Qt, OpenStack, LibreOffice) mined from Gerrit and Rietveld. In our case, on a larger dataset, RevFinder had a top-10 accuracy of 87.39\% (51 projects). However---by findings of the current paper---this dataset also included GitHub projects, thus better recommendation results. If we look into RevFinder just in 14 Gerrit projects, we get a top-10 accuracy of 79.40\%--- very similar to the one in the original paper, while RevFinder+ (considering sub-project information) reached 82.17\%.

CoreDevRec \cite{CoreDevRec} tested on five GitHub projects provides a top-5 average accuracy of 87.82\%, in our case RevFinder gets 83.08\% on 37 GitHub projects. The CORRECT approach \cite{Correct} is reported on an industrial project with top-5 accuracy higher than 90\%. However, tests on six GitHub projects report an accuracy of 85.20\%, not far from 83.08\% reported.

The cHRev approach based on building an expertise model on historical past code changes \cite{ref:zanjani2016automatically} was tested on Android (MRR=0.65), Eclipse (MRR=0.63), Mylyn (MRR=0.72), and a private dataset (MRR=0.70). In our case, Android (MRR\textsubscript{NB+}=0.68) was mined from a different period.

The RevRec hybrid approach was tested on Android (MRR=0.69), OpenStack (MRR=0.63), and Qt (MRR=0.54) projects \cite{ref:ouni2016search}. These projects were also included in our dataset, and while authors report large improvements over RevFinder, in our case NB+ tested on the same projects yielded similar results: Android (MRR\textsubscript{NB+}=0.68), OpenStack (MRR\textsubscript{NB+}=0.57), and Qt (MRR\textsubscript{NB+}=0.54).

It is challenging to compare the results in other studies. Jeong et al. \cite{ImprovingCodeReviews} report  51\%-80\% top-5 accuracy on Firefox and Mozilla Core projects, however, these constitute a small sample. Yu et al. \cite{CommenNetwork} use precision, recall and f-measure for different top-k levels, making difficult the comparison. Finally, the ReviewBot approach based on the familiarity of developers about lines of code developed/reviewed \cite{ReviewBot}, was applied experimentally to two proprietary projects (top-5=80.85\%, top-5=92.31\%) making direct comparison difficult. However, as reported in  RevFinder comparison, ReviewBot was found to provide sub-optimal performance \cite{RevFinder} and is generally considered a more limited approach in terms of results \cite{CoreDevRec}. 

\section{Conclusions}

Software code reviews are a key part of modern development processes, to increase the overall software quality and reduce overall costs \cite{ref:kemerer2009impact, Factors_Influencing_Reviews}. The main issue in large distributed systems development is to find the appropriate code reviewers, as idle time / wrong assignments can be costly for projects \cite{RevFinder}.
In this paper, we proposed a large-scale study to examine the performance of different source code reviewer recommendation algorithms (RevFinder, Na\"ive Bayes-based) to identify the best code reviewers for opened pull requests. We mined and compared data from Github and Gerrit repositories, building a large dataset of 51 projects, with more than 293K pull requests analyzed, 180K owners and 157K reviewers. 

Overall, we found that there is generally no one of the two algorithms implemented that performs better in reviewers recommendation for all projects. However, a key finding is that the performance of the models is very different on the two types of repositories analyzed (Gerrit, GitHub). Evaluation by manual inspection of the pull requests showed that difference might be due to the characteristics of Gerrit and GitHub communities: while Gerrit is a more specialized community, GitHub is more based on reviewers posting one-time source code reviews. On Gerrit each reviewer has on the average a much higher number of pull requests he/she contributed to. Another finding is that exploiting additional sub-project details allowed to improve the recommendation results, bringing statistically significant results for the Na\"ive Bayes-based approach. However, such information is only available for Gerrit projects, making difficult to have a generalized approach for all repository types.

\section*{Acknowledgment}
The work was supported from European Regional Development Fund Project CERIT Scientific Cloud (No. CZ.02.1.01/0.0/0.0/16\_013/0001802). Access to the CERIT-SC computing and storage facilities provided by the CERIT-SC Center, provided under the programme "Projects of Large Research, Development, and Innovations Infrastructures" (CERIT Scientific Cloud LM2015085), is greatly appreciated.








\newpage

\begin{table*}[]
\centering
\caption{Appendix: Information about mined projects: GitHub (\#1-\#37), Gerrit (\#38-\#51)}
\label{tbl:appendix}
\begin{tabular}{llllrrrrr}
\toprule
\# & Project            & From       & To         & Pull Reqs & Reviewers & Owners & Subprj & Files   \\
\midrule
1  & angular            & 2014-09-26 & 2017-12-08 & 5500          & 682       & 727    & 1           & 14,168  \\
2  & angular.js         & 2010-09-08 & 2017-11-02 & 6078          & 1283      & 2581   & 1           & 3,004   \\
3  & atom               & 2012-03-06 & 2017-12-07 & 2244          & 658       & 566    & 1           & 3,702   \\
4  & bitcoin            & 2011-02-21 & 2017-12-15 & 7347          & 674       & 808    & 1           & 3,629   \\
5  & bootstrap          & 2011-08-19 & 2017-12-08 & 6317          & 1949      & 2970   & 1           & 4,138   \\
6  & brackets           & 2011-12-14 & 2017-12-07 & 4264          & 260       & 507    & 1           & 4,916   \\
7  & cgm-remote-monitor & 2014-06-07 & 2017-12-03 & 1442          & 88        & 531    & 1           & 491     \\
8  & django             & 2012-04-28 & 2017-12-08 & 6912          & 723       & 2248   & 1           & 5,815   \\
9  & homebrew-core      & 2016-05-13 & 2017-12-10 & 10013         & 937       & 2364   & 1           & 4,550   \\
10 & ionic              & 2013-11-16 & 2017-11-29 & 1248          & 417       & 711    & 1           & 1,985   \\
11 & jekyll             & 2010-09-01 & 2017-12-08 & 2498          & 600       & 904    & 1           & 1,397   \\
12 & jquery             & 2010-09-05 & 2017-10-27 & 1760          & 215       & 746    & 1           & 557     \\
13 & kubernetes         & 2014-06-07 & 2017-12-10 & 29807         & 1533      & 1980   & 1           & 47,490  \\
14 & laravel            & 2011-06-10 & 2017-12-06 & 1827          & 762       & 1014   & 1           & 1,475   \\
15 & material-ui        & 2014-10-18 & 2017-12-08 & 3196          & 530       & 1122   & 1           & 6,625   \\
16 & meteor             & 2012-04-11 & 2017-12-05 & 1848          & 572       & 650    & 1           & 3,692   \\
17 & moby               & 2016-01-29 & 2017-12-12 & 7026          & 736       & 984    & 1           & 10,156  \\
18 & moment             & 2011-06-16 & 2017-11-28 & 1182          & 358       & 759    & 1           & 1,061   \\
19 & oh-my-zsh          & 2010-08-31 & 2017-12-06 & 2224          & 1234      & 1638   & 1           & 1,321   \\
20 & opencv             & 2012-07-26 & 2017-12-13 & 6309          & 386       & 1136   & 1           & 9,887   \\
21 & react              & 2013-05-29 & 2017-11-11 & 5048          & 945       & 1735   & 1           & 5,572   \\
22 & react-native       & 2015-01-30 & 2017-11-09 & 5014          & 1261      & 2043   & 1           & 5,341   \\
23 & redis              & 2010-10-03 & 2017-12-13 & 840           & 222       & 423    & 1           & 1,535   \\
24 & redux              & 2015-06-02 & 2017-12-05 & 1109          & 317       & 737    & 1           & 1,245   \\
25 & requests           & 2011-02-19 & 2017-11-28 & 1542          & 272       & 763    & 1           & 617     \\
26 & revealjs           & 2011-12-21 & 2017-12-02 & 431           & 103       & 338    & 1           & 802     \\
27 & rxjava             & 2013-01-23 & 2017-12-06 & 2353          & 148       & 313    & 1           & 5,853   \\
28 & scikit-learn       & 2010-09-01 & 2017-12-15 & 4809          & 427       & 1294   & 1           & 2,495   \\
29 & spark              & 2016-04-20 & 2017-12-13 & 7007          & 552       & 869    & 1           & 6,837   \\
30 & spring-boot        & 2013-05-22 & 2017-12-11 & 1676          & 212       & 634    & 1           & 5,068   \\
31 & spring-framework   & 2011-02-09 & 2017-12-09 & 1056          & 85        & 502    & 1           & 5,364   \\
32 & swift              & 2015-11-03 & 2017-12-09 & 7255          & 527       & 581    & 1           & 11,351  \\
33 & tensorflow         & 2015-11-09 & 2017-12-09 & 4980          & 590       & 1608   & 1           & 13,029  \\
34 & threejs            & 2010-09-03 & 2017-12-07 & 4802          & 422       & 1212   & 1           & 5,122   \\
35 & vue                & 2014-02-04 & 2017-12-06 & 658           & 156       & 325    & 1           & 679     \\
36 & webpack            & 2012-05-01 & 2017-12-08 & 1178          & 295       & 528    & 1           & 2,288   \\
37 & yarn               & 2016-02-04 & 2017-12-08 & 1153          & 468       & 416    & 1           & 4,667   \\
\midrule
38 & android            & 2008-10-24 & 2012-01-26 & 5029          & 93        & 346    & 111         & 26,768  \\
39 & chromium           & 2011-05-05 & 2017-11-04 & 3225          & 631       & 766    & 72          & 23,429  \\
40 & eclipse            & 2012-02-10 & 2017-10-29 & 9159          & 315       & 436    & 222         & 108,366 \\
41 & gem5               & 2017-01-10 & 2017-11-02 & 344           & 30        & 27     & 4           & 1,774   \\
42 & gerrit             & 2012-10-25 & 2017-12-06 & 3156          & 90        & 89     & 75          & 6,178   \\
43 & go                 & 2015-05-15 & 2017-10-28 & 4895          & 131       & 518    & 29          & 7,113   \\
44 & gwt                & 2012-10-13 & 2017-10-27 & 2870          & 47        & 153    & 8           & 13,415  \\
45 & kitware            & 2010-08-25 & 2017-11-03 & 11493         & 210       & 286    & 23          & 49,467  \\
46 & libreoffice        & 2013-05-05 & 2017-12-06 & 5870          & 90        & 241    & 21          & 19,156  \\
47 & lineageos          & 2015-05-16 & 2017-11-04 & 10977         & 256       & 241    & 558         & 14,018  \\
48 & openstack          & 2011-07-18 & 2012-05-30 & 6545          & 82        & 324    & 35          & 11,409  \\
49 & qt                 & 2011-05-17 & 2012-05-25 & 23665         & 200       & 444    & 57          & 77,767  \\
50 & scilab             & 2010-04-07 & 2017-12-04 & 14614         & 36        & 65     & 2           & 58,690  \\
51 & typo3              & 2011-02-23 & 2017-11-03 & 31582         & 651       & 622    & 177         & 61,118 \\


\bottomrule
\end{tabular}
\end{table*}

\end{document}